**Is the Binding Energy of Galaxies related to their Core Black Hole Mass?**

C. Sivaram and Kenath Arun

Indian Institute of Astrophysics, Bangalore

**Abstract:** Most of the large galaxies host a supermassive black hole, but their origin is still not well understood. In this paper we look at a possible connection between the gravitational binding energies of large galaxies etc. and the masses of their central black holes. Using this relation (between gravitational binding energy of the host structure and the black hole energy) we argue why globular clusters are unlikely to harbour large black holes and why dwarf galaxies, if they have to host black holes, should have observed mass to light ratios of ~100.

It is by now well established that most large galaxies (spiral, elliptical, etc.) host a super-massive black hole in their centre [1]. Again AGN's, quasars, etc. are powered by the gravitational energy of matter accreting on to these central supermassive black holes [2]. These black hole masses are typically of several millions of solar masses and can be as large as a few billion solar masses (like in the case of M87).

The origin of these supermassive black holes at the galactic cores is still an enigma. Again the black hole masses have been related to the galactic bulge mass, suggesting a common origin [3]. Many other large stellar conglomerations like globular clusters or dwarf galaxies also do not seem to host massive black holes in their core regions.

Here we propose a possible connection between the gravitational binding energies of large galaxies etc. and the masses of their central black holes. The idea is similar to what happens in core collapse of massive stars. The gravitational binding energy released in stellar collapse is carried away by neutrinos mainly (like in the case of SN1987A). The total energy carried away is just the gravitational binding energy of the remnant neutron star. [4, 5]

In the case of SN1987A, the estimated $3\times10^{53}$ ergs carried away by neutrinos, just corresponds to the binding energy of a 1.4 $M_{\odot}$ neutron star [5]. More massive stars would release more binding energy and this would correspond to the formation of a black hole. The gravitational binding energy released in the formation of a black hole would correspond to its rest energy, i.e. to the mass of the black hole. (For the same reason, a supernova would not give rise to a white dwarf as a remnant, as the binding energy of the white dwarf is two or more orders smaller than the energy released in the explosion.)

By analogy, we argue that whatever processes led to the formation of the black holes at the galaxy cores was connected with the formation of the galaxy and should thus be related to the gravitational binding energy of the final structure which forms.

We note that in case of galactic structures, the binding energy of the parent galaxy is comparable to that of the black hole hosted by the galaxy. The binding energy of a typical galaxy like the Milky Way is given by:

$$(G.B.E)_{spiral} = \frac{GM_{spiral}^2}{R_{spiral}} \approx 10^{61}\, ergs \qquad \text{... (1)}$$

$( M_{spiral} \approx 10^{12} M_{\odot}, R_{spiral} \approx 10 kpc)$

The galaxy harbours a $\sim 10^6 M_{\odot}$ black hole. The energy (mass) of the black hole is:

$$(B.E)_{BH} = M_{BH} c^2 = \frac{GM_{BH}^2}{R_{BH}} \approx 10^{61}\, ergs \qquad \text{... (2)}$$

The equality of equations (1) and (2) is suggestive of what happens in the release of gravitational binding energy in stellar collapses leading to the formation of remnant compact object.

Again for large elliptical galaxies, mass is few orders higher, that is: $M_{elliptical} \approx 10^{14} M_{\odot}$, so the binding energy is:

$$(B.E)_{elliptical} \approx 10^{63}\, ergs \qquad \text{... (3)}$$

So galaxies like M87 harbour billion solar mass black hole, whose binding energy:

$$(B.E)_{BH} \approx 10^{63}\, ergs \qquad \text{... (4)}$$

Therefore we have in the case of the galaxies; the binding energy of the galaxy being comparable to the black hole energy:

$$M_{BH} c^2 = \frac{GM_{galaxy}^2}{R} \qquad \text{... (5)}$$

Hence the fraction of black hole mass of the galactic mass is given by:

$$\frac{M_{BH}}{M_{galaxy}} = \frac{GM_{galaxy}}{Rc^2} \approx 10^{-6} \quad \text{... (6)}$$

This implies that smaller galaxies of $10^8 M_\odot$, could harbour black holes of $100 M_\odot$.

This could also provide a reason as to why globular clusters do not host a black hole. The binding energy of the cluster is:

$$(B.E)_{cluster} = \frac{GM^2_{cluster}}{R_{cluster}} \approx 10^{51}\, ergs \quad \text{... (7)}$$

which is lesser than the binding energy of even neutron stars.

In the case of some globular clusters like the G2 in the Andromeda galaxy, there is evidence of a $\sim 10^3 M_\odot$ black hole. This could have been formed at an earlier epoch when the cluster was more closely packed or due to merger of few clusters.

The energy of the $10^3 M_\odot$ black hole is: $E = M_{BH}c^2 \approx 10^{57}$ ergs

In order for the cluster to have a binding energy of this order, its size should be:

$$R = \frac{GM^2_{cluster}}{M_{BH}c^2} \sim 10^{-4}\, pc \quad \text{... (8)}$$

Where, $M_{cluster} \sim 10^6 M_\odot$

As the potential decreases, the kinetic energy of the cluster increases, hence spreading it out over time.

Similarly certain dwarf galaxies could also harbour $10^3 M_\odot$ black hole, even though their binding energy (with visible mass) suggests that they shouldn't.

The binding energy of the dwarf galaxy is:

$$(B.E)_{dwarf} = \frac{GM^2_{dwarf}}{R_{dwarf}} \approx 10^{55}\, ergs \quad \text{... (9)}$$

And the energy of the $10^3 M_\odot$ black hole is of the order of $10^{57}$ ergs.

This would suggest that there is substantial amount of dark matter to make up for the binding energy of the dwarf galaxy, that is: [6]

$$\frac{M}{L} \approx 100 \quad \text{... (10)}$$

References:

1. L. Kaper *et al*, Black Hole in Binaries and Galactic Nuclei, Springer, N.Y, 2001
2. F. Melia, The Galactic Supermassive Black Hole, Princeton N.J, PUP, 2007
3. Ya. B. Zel'dovich and I. D. Novikov, Relativistic astrophysics part I, Pergamon, 1971; S. Shapiro and S. Teukolsky, White Dwarf, Neutron Stars and Black Holes, Wiley, 1982
4. A. Burrows, Ann. Rev. Nucl. Part. Sci., 40, 181, 1990
5. J. Bahcall, Neutrino Astrophysics, CUP, 1989
6. S. Faber and J. Gallagher, Annual Review of Astronomy and Astrophysics, Vol. 17, pp. 135-187, 1979; E. W. Kolb and M. S. Turner, The Early Universe, Addison-Wesley, Redwood City, 1990